# Optimal Deployment of Drone Base Stations for Cellular Communication by Network-based Localization

Xiaohui Li[1], Li Xing[2]

1. School of Electrical Engineering and Telecommunications, University of New South Wales, Sydney 2052, Australia
E-mail: xiaohui.li@student.unsw.edu.au

2. Department of Electrical and Electronic Engineering, University of Melbourne, Melbourne 3010, Australia
E-mail: l.xing@student.unimelb.edu

**Abstract:** Drone base stations can assist cellular networks in a variety of scenarios. To serve the maximum number of users in an area without apriori user distribution information, we proposed a two-stage algorithm to find the optimal deployment of drone base stations. The algorithm involves UTDOA positioning, coverage control and collision avoidance. To the best of our knowledge, the concept that uses network-based localization to optimize the deployment of drone-BSs has not been analyzed in the previous literature. Simulations are presented showing that the proposed algorithm outperforms random search algorithm in terms of the maximum number of severed users under the deployment of drone-BSs they found, with limited user densities.

## 1 Introduction

Traditional terrestrial base stations (BSs) may not fully satisfy the high reliability and resilience demands of modern cellular networks in a variety of temporary or emergency events, where terrestrial infrastructures could be malfunctioned, and network traffic could be congested [1]. With recent advancements in drone technology, using drone mounted base stations (drone-BSs) in wireless cellular networks has attracted considerable attention. As a rapid solution to provide wireless connectivity, drone-BSs can assist cellular networks in cases of boosting post-disaster rehabilitation, providing connectivity in rural areas and relieving network congestion caused by excessive traffic [2]. Unlike terrestrial BSs, the locations of drone-BSs are not fixed. One of the critical issues of using drone-BSs is to determine their optimal placement. Clearly, drone-BSs should be deployed at the locations where the maximum number of users can be covered.

A growing number of papers have been published on the 2D and 3D placement of drone-BSs in cellular networks. [3] proposed a polynomial-time successive algorithm to address the placement problem of drone-BSs. [1] proposed a computationally efficient numerical algorithm to solve the 3D placement problem of drone-BSs, based on the interior-point optimizer and bisection search. [4] proposed a heuristic algorithm to find the 3D placement of drone-BSs in an area with different user densities. [5] optimized the locations of drone-BSs by an exhaustive search algorithm, aims to maximize the number of covered users using the minimum transmit power. [6] proposed an optimal placement algorithm that utilizes brute force search technique to improve the throughput and coverage of the network. A common limitation of previous literature, however, is that they assumed the accurate locations of users in the operation area are apriori. In practice, network users' location information is often unavailable. For users that are not covered by any functional BS, their exact locations cannot be acquired by the cellular network. [7] formulated a series of single and multiple drone-BSs deployment problems and proved that they are NP-hard. Several greedy algorithms were designed to solve the proposed problems. The paper only considers deploying drone-BSs in urban area that can be described by a street graph with apriori MS density function that showing the traffic demand at a certain position on the street. Besides, the 2D projections of drone-BSs are restricted to streets to avoid collision with buildings, and MSs to be served are assumed to be located near streets. [7] innovatively considered drone-BSs' battery constraints and inner drone distance constraint during optimizations. In our paper, the operating area of drone-BSs is not limited to urban areas, and MSs are not necessarily to be located near streets. Moreover, we optimize the deployment of drone-BSs with the assumption that no any apriori distribution information of MSs is available, including the MS density function.

From the second to the fourth generation, cellular networks intrinsically support network-based positioning/localization. By using measurements that are collected within the network such as time-of-arrival (TOA), time-difference-of-arrival (TDOA) and angle-of-arrival (AOA), network-based localization can determine the locations of mobile stations (MSs) without the aid of any external sources [8]. TA (Timing Advance), EOTD (Enhanced Observed Time Difference) [9], UTDOA (Uplink Time Difference of Arrival) and GPS based method are four standard localization methods for cellular networks. Practically, two dominating technologies that are commonly used for MSs' localization are GPS and UTDOA because of their high positioning accuracy. Both technologies can perform well in perfect conditions. However, GPS requires a clear view of multiple satellites. Thus, it has difficulty in locating indoor MSs and MSs in an urban area with tall buildings. Besides, GPS relies on particular chipsets while UTDOA does not require any built-in hardware features in MSs [10]. Theoretically, UTDOA can locate any type of MSs within the network, regardless of any hardware modification or software updates. Thus, UTDOA is the only technology that is uniquely suited for mission-critical scenarios such as emergency call localization. Another advantage of UTDOA is its low impact on MSs' battery life because the main computational effort for position estimation is made by the network [11].

---

*This work was supported by the Australian Research Council.

Thus, the objective that using drone-BSs to cover the maximum number of users can be achieved by deploying grouped drone-BSs to locate users in the operating area by UTDOA positioning method, then decides their optimal placements based on detected user locations. Firstly, drone-BSs need to thoroughly explore the operating area so that every user in the area can be located by drone-BSs. In other words, drone-BSs need to achieve sweep coverage [12, 13] in the operating area. Note that the 'coverage' here is different from the coverage of cellular BSs. Regarding coverage control, [13] defined three types of coverage, namely: barrier coverage [14], blanket coverage [15] and sweep coverage [16]. For barrier coverage and blanket coverage, sensor nodes are in a static arrangement to minimize the probability of undetected incursion through the sensing barrier and maximize the detection rate of targets in a sensing area, respectively. Meanwhile, sweep coverage aims to steer a group of sensor nodes so that every point in the sensing area is detected. In this paper, we consider each drone-BS fleet as a sensor node to find the locations of all users while sweeping the entire operating area. Previous studies relating to coverage control problem of a group of mobile robots including [12, 17, 18]. [17] presented a decentralized control law for mobile robots to accomplish sweep coverage based on consensus between neighbor robots. [14] proposed the control algorithms for self-deployed mobile robots to address the problems of barrier coverage and sweep coverage, based on the nearest neighbor rule and information consensus. The coverage control algorithms proposed in [11, 13, 14] are all decentralized and based on the assumption that the mobile robots' communication ability is limited. Thus, the robots can only communicate with their neighbors. Since the drone-BSs we considered are used to provide continuous wireless connectivity. We assume that drone-BSs have full ability to communicate with the control center and all other drone-BSs. The sweep coverage problems previously studied in [19-21] are related to complete coverage path planning problems. Solutions for this type of problems usually require the map of the operating area to be apriori or able to be constructed online. A group of cooperative robots can completely cover an operating area follows the sweeping path generated by path planner or control center based on the map. To solve the sweep coverage problem of drone-BSs, we partly adopt the coverage path planning procedure proposed in [21], with the assumption that the map of the operating area is apriori. Accordingly, we proposed a centralized control algorithm for drone-BSs to achieve sweep coverage by area decomposition and zigzag sweeping path.

To achieve sweep coverage by a group of cooperative drone-BSs, collision avoidance between drone-BSs becomes a problem needs to be solved. As a fundamental problem of robotics, collision-free navigation of mobile robots has been the subject of extensive research [22]. Collision avoidance between a group of mobile robots was previously studied in [22-27]. Where [24] introduced a reactive navigation control strategy of a mobile robot in prior unknown environments with moving and deforming obstacles. [25] proposed a control law for mobile robots to avoid collisions based on sliding mode control. [22] presented a collision-free navigation algorithm for a mobile robot in unknown complex environments. [27] proposed a method of collision avoidance for UAV based on simple geometric approach. In this paper, we only discuss the collision avoidance between drone-BSs. The collisions between drone-BSs and the other obstacles such as trees and high buildings are neglected. To solve the collision avoidance problem between drone-BSs, we partly adopt the method proposed in [27] because of its universality.

In this paper, we address the deployment problem of drone-BSs in the operating area with no functional terrestrial BSs, aims to serve the maximum number of users, with no apriori user distribution information. Specifically, we introduced a two-stage algorithm that performing UTDOA positioning by fleets of drone-BSs to find estimated locations of MSs, while achieving sweep coverage in the operating area. After that, determine the optimal deployment of drone-BSs by MINLP optimization based on estimated user locations. Collision avoidance between drone-BSs is also discussed. To the best of our knowledge, the concept of UTDOA positioning by drone-BSs has not been analyzed in the previous literature. The rest of this paper is organized as follows. Section 2 presents the control model and UTDOA positioning model of drone-BSs. In section 3, we introduce our two-stage algorithm. Some simulation results and discussion are presented in section 4. Finally, section 5 concludes the paper.

## 2 System Model

We consider a geographical area with $n$ uncovered MSs, denoted by the set $U = \{U_1, U_2, \cdots, U_n\}$. Where all existing terrestrial cellular BSs were manufactured, or no terrestrial BS was established. Assume that users within the operating area are either stationary or of low-mobility. A team of $m$ drone-BSs is deployed to serve as many MSs as possible in the area. According to [28], for a specific environment (urban, suburban) and a given path loss threshold, the optimal altitude and the corresponding coverage radius of the drone-BS can be numerically solved because of the following nontrivial trade-off: decreasing free space path loss requires lower altitude of the drone-BS, which in turn reducing the possibility of having line-of-sight (LoS) links for MSs. Thus, in this paper, we neglect the altitude optimization of drone-BSs and only focus on optimizing their 2D deployment.

As a network-based localization technique, UTDOA estimates MS' position by time difference among uplink signals sent from MS to a set of several BSs. The basic procedure of UTDOA positioning consists of the following four steps:

i. As a part of the serving BS, Serving Mobile Location Centre (SMLC) requests the position of the MS;
ii. MS replies by transmitting a reference signal called Sounding Reference Signal (SRS);
iii. SMLC estimates the time of arrival of SRS, and then computes the arriving time difference relating to

SRSs received by the other BSs, while the serving BS can be used as a reference;

iv. Finally, estimate the MS' position by solving the resulting non-linear equations.

The above process requires the MS to be within at least three BSs' overlapped coverage area, including the serving BS. The procedure is as illustrated in Fig. 1.

According to [11], UTDOA positioning can typically achieve a horizontal error below 50 meters in 50% of the estimations, and below 100 meters in 90% of the estimations. Moreover, the error is decreasing with the advancement of cellular technologies. Supporting vehicle-based users with high mobility has been considered throughout the design of modern cellular networks [29]. Thus, the additional error caused by drone-BSs' movement can be neglected or compensated by preventing the speed of drone-BS from too high. With the presence of position estimating error, the estimated position of an MS by UTDOA positioning will be disk-like with a radius of $r_e$, rather than an accurate coordinate.

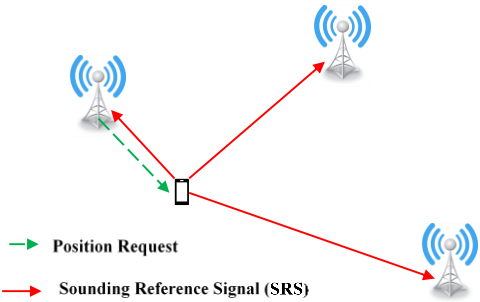

Fig. 1: Requesting SRS transmission of UTDOA

The drone-BSs are grouped into $f$ fleets, with each fleet consists of 3 drone-BSs. It can be easily seen that

$$f = m \% 3 \qquad (1)$$

By fixed formation control, the 3 drone-BSs $D = \{D_1, D_2, D_3\}$ in a fleet will be at the same fixed altitude with the mutual distance between any two of them equals to $d$. The coverage of a drone-BS is disk-like with a coverage radius of $R_C$ and the drone-BS' projection on the ground as the center, as depicted in Fig.2.

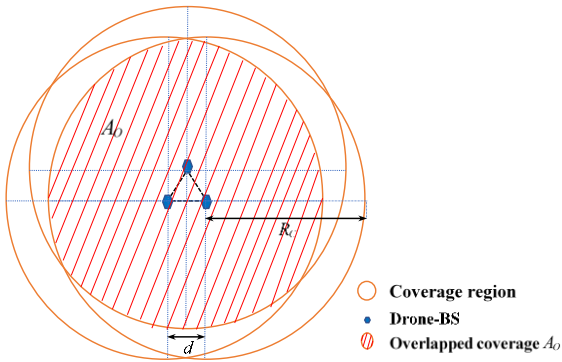

Fig. 2: Overlapped coverage of drone-BSs

By simple geometry calculation, the drone-BS fleet's overlapped coverage area $A_O$ can be formulated as

$$A_O(d, R_c) = R_c^2 (3\alpha - \frac{\pi}{2}) - \frac{3d}{2}\sqrt{R_c^2 - \frac{d^2}{4}} + \frac{\sqrt{3}}{4}d^2, \quad \alpha = \arccos(\frac{d}{2R_c}) \qquad (2)$$

As shown in equation (2), the overlapped coverage area $A_O$ is determined by drone-BSs' mutual distance $d$ and the coverage radius $R_C$. Noting that $d$ must be carefully set because of the following trade-off: decreasing $d$ will lead to larger overlapped coverage area, benefits the sweep coverage and reduces the entire mission time, which in turn reduces the positioning accuracy because the arriving time difference will be less significant.

The operating area is modeled as a convex polygon in the Cartesian coordinate plane, with an area of $A_P$. Throughout this paper, we assume that

$$A_p \gg fA_O \qquad (3)$$

Accordingly, the drone-BS fleets need to efficiently explore a relatively large operating area to find the optimal deployment. During the flight, the fleets will request positions of all MSs within the overlapped coverage area. Assuming that MSs are sparsely distributed in the operating area with a density of $\lambda_u$.

## 3 Proposed Algorithm

We now describe how to find the optimal deployment for drone-BSs to serve the maximum number of users by a two-stage algorithm. The first stage is dealing with sweep coverage control and MSs positioning by UTDOA method, and the second stage finds the optimal 2-D deployment of drone-BSs by MINLP optimization. Collision avoidance between drone-BSs is also introduced.

### 3.1 Sweep coverage control and MSs positioning

In the first stage, drone-BS fleets need to thoroughly explore the operating area to estimate the positions of all MSs inside, i.e., to achieve sweep coverage in the operating area[12, 17]. To efficiently achieve sweep coverage, we first decompose the operating area into several sub-areas and ensure that each sub-area is assigned to a fleet of three drone-BSs. After that, deploy each drone-BS fleet to its assigned sub-area, sweep the area in the optimal direction follows a zigzag pattern until 100% sweep coverage is achieved in the operating area. The optimal direction can minimize the number of turns needed along the zigzag pattern, thus reducing the total mission time. The zigzag sweeping paths are generated by the control center. The goal is to let every MS pass through the overlapped coverage to estimate positions of all MSs. When the drone-BS fleet flies over a sub-area, the position of each MS within the overlapped coverage will be requested by one of the drone-BSs that is closest to the MS. The position will then be recorded by the control center for following optimal deployment calculation.

We adopt the area decomposition algorithm proposed in [21] because of its simplicity. The inputs of the algorithm are the vertexes of the convex polygon $P$, the number of sub-areas $s$ and the proportions $\{p_1, p_2 \cdots p_s\}$ of the area that should be assigned to each sub-area. Where $s$ equals to the number of available drone-BS fleet $f$. The outputs of the algorithm are a set of sub-areas $\{S_1, S_2 \cdots S_s\}$ and the optimal sweep direction. The proportions $\{p_1, p_2 \cdots p_s\}$ can be determined by the relative capabilities of the drone-BS fleets, such as remaining battery life. The summarized area decomposition algorithm can be seen in Table 1, and more details can be seen in [21]. Examples of the polygon area decomposition with two available drone-BS fleets and equal proportions can be seen in Fig. 3, where the blue arrows stand for the zigzag sweeping path of the fleets. When finishing the first stage, the estimated positions of all MSs within the operating area will be obtained.

### 3.2 Find the optimal 2-D deployment

In the second stage, we determine the optimal 2-D deployment based on the obtained MSs' position estimations, aims to cover as many MSs as possible with a limited number of drone-BSs. We partly adopt the algorithm proposed in [1]. As discussed in section 2, the estimated position of an MS is disk-like with a radius of $r_e$, which equals to the position estimating error. The actual position of the MS could be at any point on the position estimation disk. Our goal can be achieved by placing drone-BSs' coverage area to cover as many position estimation disks as possible. Let the set $C = \{C_1, C_2, \cdots, C_n\}$ denotes all the position estimation disks obtained by the first stage of our algorithm. The center locations of $C$ are denoted by $\{(x_1, y_1), (x_2, y_2), \cdots (x_n, y_n)\}$.

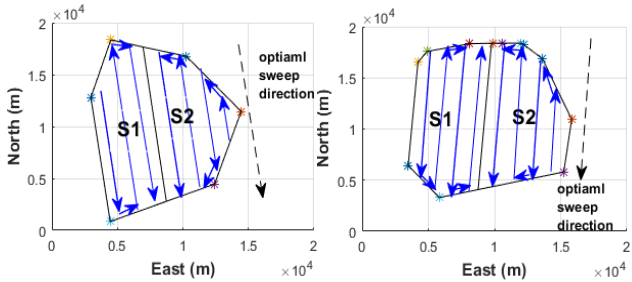

Fig. 3: Polygon area decomposition examples

Table 1: Area decomposition algorithm

**Algorithm 1:** Area Decomposition

1: Create the diameter function of the original polygon.
2: Find the optimal sweep direction that is vertical to the direction that gives minimum diameter.
3: Slice the original polygon with divide lines that are parallel with the optimal sweep direction to match the chosen sub-area proportions.

As discussed in section 2, the coverage area of a drone-BS is also disk-like with a coverage radius of $R_C$. Let set $\{(x_D^1, y_D^1), (x_D^2, y_D^2), \cdots (x_D^m, y_D^m)\}$ be the center locations of all available drone-BSs. Noting that the drone-BSs are no longer in fleets after accomplishing sweep coverage. Let $u = \{u_1, u_2, \cdots u_n\}$ be a set of binary decision variables such that $u_i \in \{0,1\}$, $i \in C$. Let $u_i = 1$ when the position estimation disk $C_i$ is entirely covered by one of the drone-BSs with the central location $(x_D^j, y_D^j)$, and $u_i = 0$ otherwise. $C_i$ is entirely covered if its center is located within a distance of $(R_C - r_e)$ from $(x_D^j, y_D^j)$. This constraint can be formulated as

$$u_i((x_i - x_D^j)^2 + (y_i - y_D^j)^2) \leq (R_C - r_e)^2 \qquad (4)$$

To ensure that equation (4) can be satisfied when $u_i = 0$, we further rewrite equation (4) as

$$(x_i - x_D^j)^2 + (y_i - y_D^j)^2 \leq (R_C - r_e)^2 + P(1 - u_i) \qquad (5)$$

Where $P$ is a large constant that satisfies equation (5) when $u_i = 0$. It can be seen that equation (5) reduces to equation (4) when $u_i = 1$. Firstly, we consider deploying one drone-BS to cover the maximum number of position estimation disks. The 2-D deployment problem is then formulated as

$$\begin{aligned}
&\underset{x_D^j, y_D^j, u_i}{\text{maximize}} \sum_{i \in C} u_i \\
&\text{subject to} \\
&(x_i - x_D^j)^2 + (y_i - y_D^j)^2 \leq (R_C - r_e)^2 + P(1 - u_i), \forall i \in C \\
&u_i \in \{0,1\}, \forall i \in C
\end{aligned} \qquad (6)$$

Let $C^{\text{cov}} \subseteq C$ denotes the set of covered users obtained by solving (6). The problem (6) is a mixed integer non-linear problem (MINLP), which can be solved to find $x_D^j, y_D^j, C^{\text{cov}}$ by interior-point optimizer of MOSEK solver [1]. Subsequently, let $C = C - C^{\text{cov}}$ and solve (6) again to find the optimal placement of another drone-BS. Then repeat the procedure until found the optimal deployment for all drone-BSs. The algorithm to find the optimal 2-D deployment is as summarized in Table 2.

Table 2: Optimal 2-D Deployment algorithm

**Algorithm 2:** Find the Optimal 2-D Deployment

**Input:** $\{(x_1, y_1), (x_2, y_2), \cdots (x_n, y_n)\}$, $R_C$, $r_e$

**Output:** $\{(x_D^1, y_D^1), (x_D^2, y_D^2), \cdots (x_D^m, y_D^m)\}$

1. For $j = 1 : m$
2.    Obtain $(x_D^j, y_D^j)$ and $C^{\text{cov}}$ by solving (6)
3.    $C = C - C^{\text{cov}}$
4. end

### 3.3 Collision avoidance

As discussed previously, we focus on avoiding collisions between drone-BSs. When achieving sweep coverage, each drone-BS fleet is operating in its own assigned sub-area. Since the sub-areas generated by area decomposition algorithm do not overlap with each other, the fleets will not collide with others during achieving sweep coverage. The collisions between drone-BSs may occur when they are flying between ground drone bases and the operating area. To avoid collisions, drone-BSs are configured to share their locations and velocities to all other drone-BSs through the control center. With the presence of transmitting delay and measuring error, the shared locations of drone-BSs are disk-like with a radius of $r_s$, rather than an accurate coordinate.

We now partly adopt the method proposed in [27] to achieve collision avoidance. To avoid collisions between two drone-BSs, the minimum distance between them should be larger than a specified minimum separation distance $d_{safe}$ when they are passing each other. Naturally, we have:

$$d_{safe} > 2r_s \qquad (7)$$

The pass distance can be found by Point of Closest Approach [27]. As illustrated in Fig.4, the pass distance vector $\vec{l}_p$ is defined as:

$$\vec{l}_p = \hat{c} \times (\vec{l} \times \hat{c}) \qquad (8)$$

Where $\vec{l}$ is the relative distance vector and $\hat{c}$ is the unit vector in the direction of the relative velocity vector $\vec{c}$ from drone-BS 'A' to drone-BS 'B'. The magnitude of $\vec{l}_p$ is the pass distance $\|\vec{l}_p\|$. It can be seen that the pass distance vector $\vec{l}_p$ and the relative velocity vector $\vec{c}$ are orthogonal:

$$\vec{l}_p \cdot \vec{c} = 0 \qquad (9)$$

The time to the closet approach $\tau$ satisfies:
$$\vec{l}_p = \vec{l} + \vec{c} \cdot \tau \quad (10)$$
By combining (9) and (10), we can get:
$$\tau = \frac{\vec{l} \cdot \vec{c}}{\vec{c} \cdot \vec{c}} \quad (11)$$

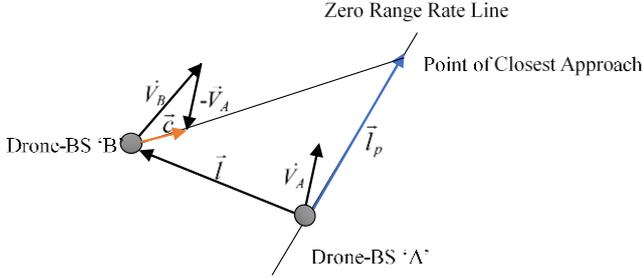

Fig. 4: Relative movement of two drone-BSs

To avoid collisions between two drone-BSs, we define $l_{\arg in}$ as the margin pass distance. It can be seen that
$$l_{m\arg in} = \|\vec{l}_p\| - d_{safe} > 0 \quad (12)$$
is the condition of no collision. If $l_{m\arg in} < 0$ was found, the control input $\vec{U}_A$ and $\vec{U}_B$ should be taken to increase $l_{m\arg in}$ by driving two drone-BSs to the opposite directions in parallel with $\vec{l}_p$, as shown in Fig. 5. Where $\vec{l}_{VSA}$ and $\vec{l}_{VSB}$ are the vectors on the directions of $\vec{l}_p$ and $-\vec{l}_p$, respectively. The increased margin pass distance $l^*_{m\arg in}$ becomes:
$$l^*_{m\arg in} = \|\vec{l}_p\| + \|\vec{l}_{VSA}\| + \|\vec{l}_{VSB}\| - d_{safe} \quad (13)$$

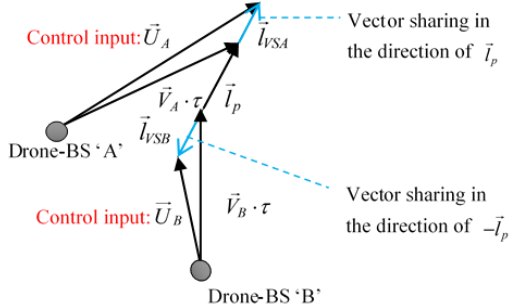

Fig. 5: Control input illustration

By setting $l^*_{m\arg in}$ to a predefined value that is larger than zero, $\vec{l}_{VSA}$ and $\vec{l}_{VSB}$ can be obtained by:
$$\vec{l}_{VSA} = \frac{l^*_{m\arg in} \cdot |\vec{V}_B|}{|\vec{V}_A| + |\vec{V}_B|} \left( \frac{\vec{l}_p}{|\vec{l}_p|} \right) \quad (14)$$
$$\vec{l}_{VSB} = \frac{l^*_{m\arg in} \cdot |\vec{V}_A|}{|\vec{V}_A| + |\vec{V}_B|} \left( \frac{\vec{l}_p}{|\vec{l}_p|} \right) \quad (15)$$
Where $\vec{V}_A$ and $\vec{V}_B$ are the velocities of drone-BS 'A' and 'B'. Finally, the control input $\vec{U}_A$ and $\vec{U}_B$ can be calculated by:
$$\vec{U}_A = \vec{V}_A \cdot \tau + \vec{l}_{VSA} \quad (16)$$
$$\vec{U}_B = \vec{V}_B \cdot \tau + \vec{l}_{VSB} \quad (17)$$
$$\vec{U}_A, \vec{U}_B \in \vec{U}_l \quad (18)$$
Where $\vec{U}_l$ sets the control limit. The collision between drone-BSs can be avoided by applying control input $\vec{U}_A$ and $\vec{U}_B$ at the time point when condition (12) was found to be not holding.

## 4 Simulation Results

To assess the performance of the proposed algorithm, we simulated the deployment of a team of 6 drone-BSs in a 10km×10km quadrangle area with no operational terrestrial base station. User distribution in the area is generated by independent Poisson point processes (PPPs), with the user density of $\lambda_u$. The position estimating error is $r_e$. The simulation parameters are as shown in Table 3. Drone-BSs have no apriori information on users' locations. The drone-BSs were first divided into two fleets to find the estimated locations of MSs, then 6 optimal placements for drone-BSs were found by the proposed algorithm.

Table 3: Simulation parameters

| Parameter | Value |
| --- | --- |
| Drone-BS flight speed $v$ | 10 m/s |
| Coverage radius $R_C$ | 500 m |
| Mission time | 100 minutes |
| Sub-area proportions $\{p_1, p_2 \cdots p_s\}$ | {0.5,0.5} |
| Mutual distance $d$ | 50m |

The performance of the proposed algorithm was compared to that of the random search algorithm. For random search algorithm, drones-BSs move along random initial directions. Moreover, they only change directions when their coverage areas hit other drones-BSs' coverage areas or the operating area boundaries. The number of connected users is recorded and compared by drone-BSs in real time to find the optimal deployment. Fig. 6 and Fig. 7 show how the number of severed users under the deployment of 6 drone-BSs found by different algorithms increases with user density, for $r_e$=30m and $r_e$=0, respectively. As can be seen from Fig. 6, the number of severed users under the deployment found by proposed algorithm is higher than that of the random search method when the user density falls below a certain value. The result could be attributed to the fact that the operating area with lower user density is more difficult for the random search method to find users. By comparing Fig. 6 and Fig. 7, we can find that the random search method outperforms proposed algorithm for the area with higher user density mainly because of the position estimation error. Accordingly, proposed algorithm will be more feasible when the accuracy of UTDOA positioning method increases. Besides, the estimated user positions found by proposed algorithm can also be used for other applications such as disaster rescue.

## 5 Conclusion

In this paper, the optimal deployment of drone-BSs without apriori user distribution information is investigated. The aim is to cover as many users as possible by a limited number of drone-BSs. We proposed a two-stage algorithm that utilizes drone-BS fleets to find the estimated locations of users in the operating area by UTDOA positioning method. Then we determine the optimal deployment of the drone-BSs by MINLP optimization, based on estimated user locations. Simulations have shown that the proposed algorithm outperforms random search algorithm regarding the maximum number of severed users under the deployment of drone-BSs they found, with limited user densities. In the future work, the collision avoidance between drone-BSs and other obstacles could be considered. The future study could also extend to the case with non-stationary users.

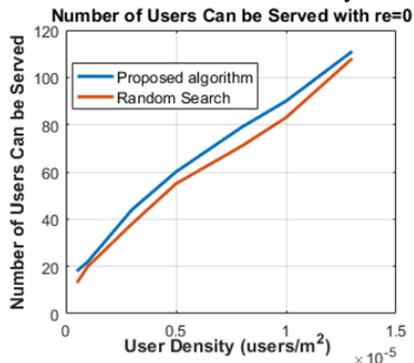

Fig. 6: Number of severed users versus user density for different algorithms, with 6 drone-BSs and $r_e=30$m

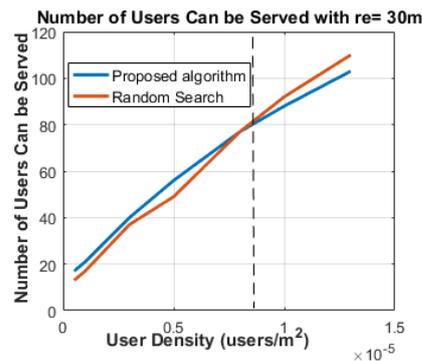

Fig. 7: Number of severed users versus user density for different algorithms, with 6 drone-BSs and $r_e=0$